\documentclass[aps,prl]{revtex4}

\usepackage{amsmath,amssymb,amsfonts,bm,graphicx,graphics,color}
\usepackage{mathtools}
\usepackage{siunitx}
\usepackage{hyperref}
\hypersetup{hidelinks}
\usepackage{enumitem}
\usepackage{microtype}
\usepackage{booktabs}
\usepackage{algorithm}
\usepackage{algpseudocode}

\newcommand{\ket}[1]{|#1\rangle}

\newcommand{\de}{\partial}
\newcommand{\sech}[1]{\textrm{sech}\left( #1\right)}

\newcommand{\diag}{\textrm{diag}}

\newcommand{\RR}{\mathrm{RR}}
\newcommand{\Tr}{\mathrm{Tr}}

\newcommand{\Id}{I}
\newcommand{\ketvac}{\ket{0}}

\newcommand{\deltaz}{\Delta z}

\begin{document}

\title{The Quantum Split-Step Fourier Algorithm for Nonlinear Optical Waveguides}

\author{Fabio Biancalana}
\affiliation{Institute of Photonics and Quantum Sciences (IPaQS), School of Engineering and Physical Sciences, Heriot-Watt University, EH14 4AS Edinburgh, UK}

\begin{abstract}
We introduce the Quantum Split-Step Fourier (QSSF) algorithm for nonlinear optical waveguides, a numerical framework that combines split-step propagation of the nonlinear Schr\"odinger equation with a commutator-preserving Bogoliubov evolution of Gaussian quantum fluctuations. The method propagates the classical mean field together with the Bogoliubov matrices $U$ and $V$, from which reduced second moments, covariance matrices, symplectic eigenvalues, and entropic measures are constructed for arbitrary spectral windows. Applied to soliton-driven resonant radiation, QSSF shows that the selected radiation band acquires a steadily increasing von Neumann entropy and a corresponding loss of purity, quantifying its entanglement with the rest of the spectrum in the lossless Gaussian setting. The analysis also reveals a surprisingly pronounced low-dimensional structure: although the radiation occupies many Fourier bins, its reduced Gaussian state is dominated by only a few Williamson modes. QSSF therefore provides a practical information-theoretic diagnostic for quantum correlations in nonlinear frequency conversion, supercontinuum generation, and multimode squeezed-light formation in ultrafast waveguide platforms.
\end{abstract}

\maketitle

\section{Introduction and motivations}
When an optical soliton propagates in a dispersive nonlinear waveguide, higher-order dispersion terms can phase-match the soliton to linear waves, producing narrow spectral bands of resonant radiation (RR), also known as dispersive waves or (in earlier terminology) soliton-induced Cherenkov radiation \cite{Agrawal,NK1995,Skryabin2010}. In fiber optics the RR frequency is commonly understood from a phase-matching condition between the soliton and the linear dispersion relation, including nonlinear phase shifts and, when relevant, Raman-induced soliton self-frequency shift \cite{BiancalanaSkryabinYulin2004,SkryabinYulin2005,SkryabinScience2003}. RR has become a standard ingredient of supercontinuum (SC) generation and ultrafast frequency conversion in photonic crystal fibers and integrated waveguides \cite{Dudley2006,Skryabin2010}.

In gas-filled hollow-core waveguides, RR can be efficiently shifted into the ultraviolet (UV) and even vacuum-UV through pressure-tunable dispersion engineering. Past and present pioneering work demonstrated tunable vacuum-UV to visible ultrafast sources based on soliton-driven RR in gas-filled Kagome and related hollow-core fibers \cite{MakTraversRussell2013,Ermolov2015}. Recent studies have clarified scaling laws and control knobs for resonant dispersive-wave emission in hollow capillary fibers, including pressure-gradient configurations \cite{BrahmsBelliTravers2020}. Related deep-UV conversion using gas-filled hollow-core photonic crystal fibers has also been demonstrated via four-wave mixing and dispersive-wave physics \cite{BelliAbdolvandTraversRussell2019}.

A central practical question for SC/RR sources is their coherence: how reproducible is the generated spectrum from shot to shot when quantum noise is present? Dudley and Coen introduced a widely used numerical and experimental framework to quantify wavelength-resolved first-order coherence via the modulus of the complex degree of coherence $|g^{(1)}_{12}(\lambda)|$, evaluated from an ensemble of generalized nonlinear Schr\"odinger equation (GNLSE) simulations seeded with quantum-limited input noise \cite{DudleyCoen2002}.

Beyond first-order coherence, broadband nonstationary light can be characterized by second-order field coherence in the sense of Wolf's coherence theory, i.e.\ by two-point field correlations such as the two-time mutual coherence function $\Gamma(t_1,t_2)=\langle E^\ast(t_1)E(t_2)\rangle$ and the two-frequency cross-spectral density $W(\omega_1,\omega_2)=\langle \tilde E^\ast(\omega_1)\tilde E(\omega_2)\rangle$, from which spectral-temporal degrees of coherence can be constructed \cite{MandelWolf}. This ``second-order coherence'' (in the sense of field-correlation) framework was applied to SC generated in nonlinear fibers by Genty et al. \cite{GentySurakkaTurunenFriberg2010} and measured experimentally by N\"{a}rhi et al. \cite{NarhiTurunenFribergGenty2016}.  It is important to distinguish these field correlation functions from intensity correlation functions in Hanbury Brown--Twiss (HBT) interferometry, which quantify photon statistics through $g^{(2)}(\tau)=\langle I(t)I(t+\tau)\rangle/\langle I(t)\rangle^2$ and reveal bunching/antibunching behaviour \cite{HanburyBrownTwiss1956,MandelWolf}.

Optical fibers constitute a mature platform for quantum nonlinear optics. Kerr solitons are known to produce squeezing and multimode quantum correlations \cite{DrummondCarter1987,LaiHaus1989}. More broadly, macroscopic quantum light and nonlinear interferometers (including fiber-based four-wave mixing) have emphasized that effective mode structure can give an excellent description of multimode squeezing and correlations \cite{ChekhovaOu2016,Lemieux2016}. These developments strongly motivate the need of a propagation-resolved, information-theoretic quantification of quantum correlations between spectral bands during RR formation.

This work introduces the Quantum Split-Step Fourier (QSSF) algorithm, which extends the established coherence/noise program into an entanglement-based, Gaussian-state numerical framework. The key observation is that linearization of the quantum NLSE around a strong classical field yields a Bogoliubov--de~Gennes (BdG) evolution that is quadratic in the field operators and therefore preserves Gaussianity when starting from Gaussian state inputs. In the lossless BdG setting with vacuum input fluctuations, the global state remains pure; consequently, the von Neumann entropy of a chosen spectral window equals its entanglement entropy with the complementary modes. The von Neumann entropy of a state defined by a density matrix $\hat{\rho}$ is defined by  $S_{\rm vN}(\hat{\rho})=-\Tr(\hat{\rho}\ln\hat{\rho})$, and is the fundamental entropy measure in quantum mechanics. For pure bipartite states, $S_{\rm vN}$ quantifies entanglement exactly via the reduced density matrix \cite{NielsenChuang,Horodecki2009,PlenioVirmani2007}. Our QSSF algorithm computes $S_{\rm vN}$ for an arbitrary user-defined RR frequency window dynamically along propagation, precisely quantifying the entanglement between RR and the rest of the spectrum, and complements it with mode-resolved diagnostics that distinguish fast Fourier transform (FFT) bin occupations from basis-invariant symplectic (Williamson) mode occupations.

\section{Governing equations and Bogoliubov map}

The classical nonlinear Schr\"odinger equation (NLSE), describing the propagation of an arbitrary input pulse along the nonlinear waveguide, is written in dimensionless units as
\begin{equation}
i\,\partial_z A(z,t)+\hat D\,A(z,t)+\gamma|A(z,t)|^2A(z,t)=0,
\label{eq:nlse}
\end{equation}
where $A(z,t)$ is the pulse envelope amplitude, $z$ is the propagation distance along the waveguide, $t$ is the retarded time, and $\hat D$ is the linear dispersion operator, defined as
\begin{equation}
\hat{D}\equiv\sum_{m\geq 2}d_m(i\de_{t})^{m}.
\label{eq:disp}
\end{equation}
We use the standard Fourier sign convention in which $\partial_t$ is represented by $-i\omega$. With this convention, the action of $\hat D$ in the frequency domain is multiplication by the symbol
\begin{equation}
D(\omega)=\sum_{m\ge 2}d_m\,\omega^m.
\label{eq:Dsymbol}
\end{equation}
Here $d_m$ are dimensionless dispersion coefficients in the normalized propagation equation. In the numerical example that we discuss later in this paper, we use $d_2=1/2$ and write $d_3\equiv\beta_3$, so that $D(\omega)=\omega^2/2+\beta_3\omega^3$. The last term of Eq.~\eqref{eq:nlse} accounts for the Kerr nonlinearity, and we keep the explicit dimensionless constant $\gamma$ for later convenience and clarity; it may be set to unity at all times in all our dimensionless calculations. Equation~\eqref{eq:nlse} is standard in nonlinear fiber optics; see Ref.~\cite{Agrawal}.

We now write Eq.~\eqref{eq:nlse} in evolution form $\partial_z A=(\mathcal L+\mathcal N)A$ with
\begin{equation}
\mathcal L A \equiv i\hat D A,\qquad \mathcal N A \equiv i\gamma |A|^2A.
\end{equation}
For a propagation step $\deltaz$, the symmetric Strang splitting (second-order operator splitting) reads \cite{Strang1968}
\begin{equation}
A(z+\deltaz)\approx e^{\frac{\deltaz}{2}\mathcal L}\,e^{\deltaz\,\mathcal N}\,e^{\frac{\deltaz}{2}\mathcal L}A(z).
\label{eq:strang}
\end{equation}
In practice, $e^{\frac{\deltaz}{2}\mathcal L}$ is applied in frequency space as multiplication by the linear propagator
\begin{equation}
P(\omega,\deltaz)\equiv \exp\!\left(iD(\omega)\frac{\deltaz}{2}\right),
\label{eq:E}
\end{equation}
where $\omega$ is the frequency detuning from the central pulse frequency, while $e^{\deltaz\mathcal N}$ is local in time. If we have a set of discrete frequencies, we would simply have $P_{k}\equiv P(\omega_{k})$, where $k$ is the integer indexing the desired spectral line. The resulting split-step Fourier method (SSFM) is widely used for NLSE/GNLSE classical propagation \cite{Agrawal}; the split-step Fourier approach originates in early numerical wave-propagation literature \cite{HardinTappert1973}.

In order to take into account quantum effects on the propagation, we now promote the field envelope to a bosonic operator $\hat A(z,t)$ obeying the quantum NLSE
\begin{equation}
i\,\partial_z \hat A+\hat D\hat A+\gamma \hat A^\dagger \hat A\hat A=0.
\label{eq:qnlse}
\end{equation}
We assume a strong mean-field classical component and make the split
\begin{equation}
\hat A(z,t)=A(z,t)+\delta\hat a(z,t),\qquad \langle \delta\hat a(z,t)\rangle=0.
\label{eq:split}
\end{equation}
The condition $\langle \delta\hat a\rangle=0$ states that the mean field is entirely captured by $A(z,t)$; it is appropriate when the soliton is prepared in a coherent state whose displacement defines $A$ (vacuum noise around a coherent amplitude). The fluctuation operators satisfy equal-$z$ canonical commutators,
\begin{equation}
[\delta\hat a(z,t),\delta\hat a^\dagger(z,t')]=\delta(t-t'),
\label{eq:comm_cont}
\end{equation}
where $\delta(\cdot)$ is the Dirac delta distribution.

Linearizing Eq.~\eqref{eq:qnlse} in $(\delta\hat a,\delta\hat a^\dagger)$ yields the BdG equation
\begin{equation}
i\,\partial_z \delta\hat a+\hat D\,\delta\hat a+2\gamma|A|^2\delta\hat a+\gamma A^2\,\delta\hat a^\dagger=0.
\label{eq:bdg}
\end{equation}
Equation~\eqref{eq:bdg} is linear in the variables $(\delta\hat a,\delta\hat a^\dagger)$; therefore, if the input fluctuations are Gaussian (in particular, vacuum), the state remains Gaussian for all $z$. We discuss the specific properties of Gaussian states in the next section.

Since the simulation is performed on a finite temporal window $T$ sampled by $N_t$ points, we also have a finite number of frequency bins $\omega_{k}=k\Delta\omega$ with $\Delta\omega\equiv 2\pi/T$, and (for even $N_t$) the standard FFT bin indexing $k=0,1,\ldots,N_t/2-1,-N_t/2,\ldots,-1$. The Nyquist half-window is $\omega_{\max}\equiv \pi/\Delta t$, where $\Delta t\equiv T/N_t$.
We use these FFT bins as an orthonormal mode basis. Let $\hat a_k$ be the bosonic annihilation operator for the FFT bin $k$ and collect everything in a vector $\hat{\bm a}=(\hat a_1,\ldots,\hat a_{N_t})^T$. The discrete commutators are
\begin{equation}
[\hat a_j,\hat a_k^\dagger]=\delta_{jk},\qquad [\hat a_j,\hat a_k]=0,
\label{eq:comm_disc}
\end{equation}
where $\delta_{jk}$ is the Kronecker delta. The BdG evolution of Eq. (\ref{eq:bdg}) induces a Bogoliubov map
\begin{equation}
\hat{\bm a}(z)=U(z)\hat{\bm a}(0)+V(z)\hat{\bm a}^\dagger(0),
\label{eq:UV}
\end{equation}
where $U(z)$ and $V(z)$ are complex $N_t\times N_t$ matrices.
Equation (\ref{eq:UV}) can also be written in a 'Nambu form' by defining the vector $\hat{\bm \xi}\equiv(\hat{\bm a},\hat{\bm a}^{\dagger})^{T}$:
\[
\hat{\bm\xi}(z)=S(z)\hat{\bm\xi}(0),\qquad
S=\begin{pmatrix}U&V\\V^*&U^*\end{pmatrix}.
\]
Demanding that the output operators obey the same commutators as the inputs for any propagation distance $z$ yields the two strong conditions on the matrices $(U,V)$:
\begin{equation}
UU^\dagger-VV^\dagger=\Id,\qquad UV^T-VU^T=0,
\label{eq:sympl}
\end{equation}
where $\Id$ is the $N_t\times N_t$ identity matrix, $^\dagger$ denotes conjugate transpose, and $^T$ denotes transpose. The QSSF algorithm will monitor numerical violations of conditions \eqref{eq:sympl} using the Frobenius norm $\|X\|_{\mathrm F}\equiv\sqrt{\sum_{jk}|X_{jk}|^2}$:
\begin{equation}
\varepsilon_1\equiv \frac{1}{N_t}\|\big(UU^\dagger-VV^\dagger-\Id\big)\|_{F},
\qquad
\varepsilon_2\equiv \frac{1}{N_t}\|\big(UV^T-VU^T\big)\|_{F}.
\label{eq:eps}
\end{equation}
Small $\varepsilon_{1,2}$ certify a physically valid commutator-preserving Gaussian evolution, and are strong dynamical indicators of the correctness of our quantum algorithm.
The two symplecticity conditions (\ref{eq:sympl}) can be written in an alternative form;
defining 
\[
J=\begin{pmatrix}I&0\\0&-I\end{pmatrix},
\]
both constraints in Eq. (\ref{eq:sympl}) become the compact condition $SJS^\dagger=J$.

A key simplification which will be exploited by our QSSF is that the Kerr BdG coupling is local in time. If, for one substep, we neglect dispersion ($\hat D=0$) and freeze $A$ to a constant value [or, numerically, freeze at each time sample $t_n$ to $A_{\rm mid}(t_n)$], the BdG generator becomes a constant $2\times 2$ matrix for each time sample:
\begin{equation}
H_n=
\begin{pmatrix}
-\alpha_n & -\mu_n\\
\mu_n^\ast & \alpha_n
\end{pmatrix},
\qquad
\alpha_n\equiv 2\gamma|A_{\rm mid}(t_n)|^2,\quad \mu_n\equiv \gamma A_{\rm mid}(t_n)^2.
\label{eq:Hlocal}
\end{equation}
The overall signs in $H_n$ and in the resulting coefficients $(u_n,v_n)$ are fixed by writing the BdG evolution in the form $i\partial_z\hat\Psi=\hat H_{\mathrm{BdG}}\hat\Psi$ with $\hat\Psi=(\delta\hat a,\delta\hat a^\dagger)^T$; alternative sign conventions in the BdG equation lead to corresponding sign changes in $(u_n,v_n)$.

The corresponding propagator is the matrix exponential $e^{-iH_n\deltaz}$, which has the Bogoliubov form
\begin{equation}
e^{-iH_n\deltaz}=
\begin{pmatrix}
u_n & v_n\\
v_n^\ast & u_n^\ast
\end{pmatrix},
\qquad
u_n=\cos(\kappa_n\deltaz)+i\alpha_n s_n,\quad v_n=i\mu_n s_n,
\label{eq:uv_closed}
\end{equation}
with $\kappa_n=\sqrt{\alpha_n^2-|\mu_n|^2}$ and $s_n=\sin(\kappa_n\deltaz)/\kappa_n$ (or $s_n=\deltaz$ when $\kappa_n=0$). We shall use formulas (\ref{eq:uv_closed}) in one of the major steps in the QSSF algorithm.

For the discretized bosonic modes $\{\hat a_k\}_{k=1}^{N_t}$ we define the normal and anomalous second moments
\begin{equation}
N_{jk}(z)\equiv \langle \hat a_j^\dagger(z)\hat a_k(z)\rangle,\qquad
M_{jk}(z)\equiv \langle \hat a_j(z)\hat a_k(z)\rangle,
\label{eq:NM}
\end{equation}
where $\langle\cdot\rangle$ denotes expectation value in the fluctuation state. These matrices are standard in quantum optics and Gaussian-state theory \cite{GardinerZoller,WallsMilburn,Weedbrook2012,Serafini2017,Adesso2014,Ferraro2005}. Their physical meaning is transparent.
The diagonal entries $N_{kk}=\langle \hat a_k^\dagger \hat a_k\rangle$ are the mean excitation (photon) numbers of the discrete modes (in our case, FFT frequency bins). The off-diagonal entries $N_{jk}$ ($j\neq k$) quantify first-order mutual coherence between modes $j$ and $k$ in the chosen basis (discrete analogues of the field-correlation functions underlying $g^{(1)}$ measures) \cite{MandelWolf,GardinerZoller}. The anomalous correlators $M_{jk}=\langle \hat a_j \hat a_k\rangle$ quantify pair (phase-sensitive) correlations. $M\neq 0$ is the hallmark of phase-sensitive Bogoliubov correlations; when sufficiently large relative to the normal fluctuations encoded in $N$, it corresponds to squeezing and nonclassical noise redistribution \cite{CavesSchumaker1985,SchumakerCaves1985,WallsMilburn}. 
In a single-mode setting, $N$ sets the overall noise level while $M$ controls squeezing and phase-space rotation; in multimode settings, $M_{jk}$ encodes two-mode squeezing/correlations between modes $j$ and $k$.

For vacuum input fluctuations (coherent-state mean field), the Bogoliubov map \eqref{eq:UV} implies \cite{Weedbrook2012,Serafini2017}
\begin{equation}
N(z)=V(z)^{*}V^{T},\qquad M(z)=U(z)V^T(z).
\label{eq:NM_UV}
\end{equation}
Thus, once $(U,V)$ are known, the second moments (and therefore all Gaussian diagnostics) follow directly.

For each mode we define quadratures
\begin{equation}
\hat x_k\equiv \frac{\hat a_k+\hat a_k^\dagger}{\sqrt2},\qquad
\hat p_k\equiv \frac{\hat a_k-\hat a_k^\dagger}{i\sqrt2},
\label{eq:quad}
\end{equation}
and, for an $n$-mode subsystem, we collect the quadratures in the 'phase space' vector $\hat{\bm R}=(\hat x_1,\ldots,\hat x_n,\hat p_1,\ldots,\hat p_n)^T$. We define the symplectic form (canonical commutation matrix)
\begin{equation}
\Omega \equiv 
\begin{pmatrix}
0 & \Id_n\\
-\Id_n & 0
\end{pmatrix},
\label{eq:Omega_def}
\end{equation}
so that the quadratures satisfy $[\hat R_j,\hat R_k]=i\,\Omega_{jk}$.

Next, we define the real symmetric covariance matrix
\begin{equation}
V_q\equiv \frac12\langle \Delta\hat{\bm R}\Delta\hat{\bm R}^T+(\Delta\hat{\bm R}\Delta\hat{\bm R}^T)^T\rangle,
\qquad \Delta\hat{\bm R}\equiv \hat{\bm R}-\langle \hat{\bm R}\rangle.
\label{eq:Vq}
\end{equation}
For the fluctuation sector considered here, $\langle \hat a_k\rangle=0$ (coherent displacement is carried by the mean field $A$), so $\langle \hat{\bm R}\rangle=0$.

The Robertson--Schr\"odinger uncertainty principle for any physical (possibly mixed) state can be written compactly as \cite{Weedbrook2012,Serafini2017}
\begin{equation}
V_q+\frac{i}{2}\Omega \succeq 0,
\label{eq:RS_uncertainty}
\end{equation}
where $V_q$ is the covariance matrix \eqref{eq:Vq} and $\succeq 0$ denotes positive semidefinite order.

Williamson's theorem states that for any real, symmetric, positive-definite covariance matrix $V_q\in\mathbb R^{2n\times 2n}$ there exists a real symplectic matrix $S\in\mathbb R^{2n\times 2n}$ such that \cite{Williamson1936,Serafini2017,Weedbrook2012}
\begin{equation}
V_q = S^T \mathcal D_{\rm W}\,S,
\qquad
\mathcal D_{\rm W} \equiv \diag(\nu_1,\ldots,\nu_n,\nu_1,\ldots,\nu_n),
\label{eq:Williamson}
\end{equation}
with the symplectic condition
\begin{equation}
S\,\Omega\,S^T=\Omega.
\label{eq:sympl_S}
\end{equation}

The numbers $\nu_k>0$ are uniquely determined by $V_q$ (up to ordering) and are called the symplectic eigenvalues. Physically, Eq.~\eqref{eq:Williamson} means that, with the convention used here, the Gaussian canonical change of variables $\hat{\bm R}\mapsto S^{-T}\hat{\bm R}$ transforms the covariance into a direct sum of $n$ independent single-mode thermal covariances, each characterized by a single variance parameter $\nu_k$ shared by its two quadratures.

The symplectic eigenvalues can be extracted from the eigenvalue moduli of the matrix $i\Omega V_q$. Indeed, using $V_q=S^T\mathcal D_{\rm W}S$ and the identity $\Omega S^T=S^{-1}\Omega$, which follows from $S\Omega S^T=\Omega$, one obtains
\begin{equation}
i\Omega V_q
=
i\Omega S^T\mathcal D_{\rm W}S
=
iS^{-1}\Omega\mathcal D_{\rm W}S.
\label{eq:similarity_step}
\end{equation}
Therefore
\begin{equation}
S(i\Omega V_q)S^{-1}=i\Omega\mathcal D_{\rm W}.
\label{eq:similarity}
\end{equation}
Hence $i\Omega V_q$ and $i\Omega\mathcal D_{\rm W}$ are similar matrices and have the same eigenvalues. Since $\mathcal D_{\rm W}$ contains identical $\nu_k$ entries in its $x$ and $p$ blocks, $i\Omega\mathcal D_{\rm W}$ decomposes into $n$ independent $2\times2$ blocks
\[
i\begin{pmatrix}0 & \nu_k\\ -\nu_k & 0\end{pmatrix},
\]
whose eigenvalues are $\pm \nu_k$. Thus the spectrum of $i\Omega V_q$ consists of pairs $\{\pm\nu_k\}_{k=1}^n$, and the desired invariants are the positive moduli $\nu_k$.

Restricting \eqref{eq:NM} to a window $\mathcal W$ of $n$ modes (for example the FFT bins of the RR we want to study) yields $(N_{\mathcal W},M_{\mathcal W})$. In the quadrature ordering used in QSSF, the covariance can be written explicitly as
\begin{equation}
V_q=
\begin{pmatrix}
B+\Re(M_{\mathcal W}) & \Im(M_{\mathcal W})+\Im(N_{\mathcal W})\\
\Im(M_{\mathcal W})-\Im(N_{\mathcal W}) & B-\Re(M_{\mathcal W})
\end{pmatrix},
\qquad
B\equiv \Re(N_{\mathcal W})+\frac12 \Id_n,
\label{eq:Vq_NM}
\end{equation}
which is the multimode generalization of the familiar single-mode identities
$\mathrm{Var}(\hat x)=\langle \hat a^\dagger \hat a\rangle+\tfrac12+\Re\langle \hat a\hat a\rangle$ and
$\mathrm{Var}(\hat p)=\langle \hat a^\dagger \hat a\rangle+\tfrac12-\Re\langle \hat a\hat a\rangle$.
Equation \eqref{eq:Vq_NM} makes the physical roles of $N$ and $M$ explicit: $N$ contributes a phase-insensitive noise background and first-order inter-mode coherence, while $M$ encodes phase-sensitive noise redistribution, squeezing, and pair correlations.

\section{Gaussian state evolution and the von Neumann entropy}

A continuous-variable quantum state is Gaussian if its characteristic function (equivalently its Wigner function) is Gaussian in phase space \cite{BraunsteinVanLoock2005,Weedbrook2012,Serafini2017,Ferraro2005}. Gaussian states are completely specified by first moments $\langle \hat{\bm R}\rangle$ and second moments $V_q$; all higher moments can be deduced from them. The vacuum state $\ketvac$ is Gaussian with $\langle \hat a_k\rangle=0$ and $N=M=0$ in \eqref{eq:NM}, corresponding to $V_q=\Id/2$ in the normalization \eqref{eq:quad} \cite{WallsMilburn,Weedbrook2012}. Because the BdG dynamics is quadratic, it maps Gaussian states to Gaussian states; thus propagating $(U,V)$ suffices to reconstruct the reduced Gaussian state of any spectral window at each $z$.

As we saw in the previous section, Williamson's theorem implies that any Gaussian covariance matrix can be symplectically diagonalized; the symplectic eigenvalues $\nu_k$ are obtained as the positive moduli of the eigenvalues of $i\Omega V_q$ \cite{Williamson1936,Serafini2017}. Physicality requires $\nu_k\ge 1/2$. Equivalently, this is the Robertson--Schr\"odinger uncertainty condition $V_q+\frac{i}{2}\Omega\succeq 0$ in Eq.~(\ref{eq:RS_uncertainty}). Define the symplectic occupation numbers
\begin{equation}
n_k\equiv \nu_k-\tfrac12\ge 0,
\label{eq:nk}
\end{equation}
which are the mean occupations of independent thermal modes in the Williamson decomposition of the reduced Gaussian state.

The von Neumann entropy of an $n$-mode Gaussian state can now be calculated using the formula \cite{Weedbrook2012,Serafini2017}
\begin{equation}
S_{\rm vN}(\rho)=\sum_{k=1}^{n}\Big[(n_k+1)\ln(n_k+1)-n_k\ln n_k\Big],
\label{eq:SvN}
\end{equation}
and the R\'enyi-2 entropy and purity are given by the expressions
\begin{equation}
S_2(\rho)=\sum_{k=1}^{n}\ln(2\nu_k),\qquad
\mathcal P=e^{-S_2}=\prod_{k=1}^{n}\frac{1}{2\nu_k}.
\label{eq:S2}
\end{equation}

Besides entropy, it is often useful to quantify how many Williamson modes contribute appreciably to the reduced Gaussian state. For this purpose we define the effective number of occupied symplectic modes as
\begin{equation}
K_{\rm eff}
=
\frac{\left(\sum_{k=1}^{n} n_k\right)^2}
     {\sum_{k=1}^{n} n_k^2},
\label{eq:Keff}
\end{equation}
where $n_k=\nu_k-\tfrac12$ are the symplectic occupation numbers introduced in Eq.~(\ref{eq:nk}).

Equation~(\ref{eq:Keff}) is the inverse participation ratio of the normalized occupation distribution and provides a basis-independent measure of the effective dimensionality of the reduced Gaussian state. If all occupation is concentrated in a single Williamson mode, then $K_{\rm eff}=1$. Conversely, if $m$ Williamson modes carry equal occupation and all others are empty, then $K_{\rm eff}=m$. More generally, $K_{\rm eff}$ quantifies how broadly the total excitation is distributed across the symplectic spectrum.

To specialize to a RR frequency window, choose a frequency window $\mathcal W$ (RR band) consisting of $n<N_t$ FFT bins. Its Hilbert space is the tensor product of $n$ bosonic Fock spaces,
\begin{equation}
\mathcal H_{\mathcal W}=\bigotimes_{k\in\mathcal W}\mathrm{span}\{\ket{0}_k,\ket{1}_k,\ket{2}_k,\ldots\},
\label{eq:Hilbert}
\end{equation}
which is infinite-dimensional. The QSSF algorithm constructs the reduced state $\rho_{\mathcal W}$ from the restricted second moments $(N_{\mathcal W},M_{\mathcal W})$ obtained by selecting rows of $(U,V)$ and applying \eqref{eq:NM_UV}. In the lossless BdG setting, the global state remains pure, so $S_{\rm vN}(\rho_{\mathcal W})$ equals the entanglement entropy between RR and its complement. Equation~\eqref{eq:SvN} provides a clear numerical way to calculate the von Neumann entropy of an arbitrary frequency window of the total spectrum, thus allowing the construction of the QSSF algorithm that we describe next.

\section{Description of the QSSF algorithm}

Having introduced the required mathematical ingredients in the previous sections, we are now in a position to describe the precise steps of the QSSF algorithm in detail. All quantities below will be given in dimensionless units, as is customary for numerical simulations.

The first step is to set the number of temporal sampling points $N_t$ (typically a power of $2$ due to FFT requirements) and the time window of the simulation $T$, so that $\Delta t\equiv T/N_t$ gives the temporal spacing in the sampling. We then set the spatial propagation step $\Delta z$ and the total propagation length $L$. The temporal window is centered at $t=0$ via $t_{n}=(-N_t/2,\ldots,N_t/2-1)\Delta t$, which gives a Nyquist frequency half-window $\omega_{\max}=\pi/\Delta t$, and a frequency vector $\omega_{k}=\Delta\omega(0,1,\ldots,N_t/2-1,-N_t/2,\ldots,-1)$, with $\Delta\omega\equiv 2\pi/T$.

For the example considered below we take $d_2=1/2$ and include only one higher-order-dispersion coefficient, denoted $\beta_3\equiv d_3$, so that $D(\omega)=\omega^{2}/2+\beta_{3}\omega^{3}$. More generally, one can choose an arbitrary dispersion function with as many higher-order-dispersion coefficients as desired. We then define the linear propagator in frequency space as $P(\omega)=\exp(iD(\omega)\Delta z/2)$, which is the relevant propagator for the linear half-step in the Strang splitting of Eq. (\ref{eq:strang}).

We then initialize the pulse with our desired input pulse; in this example we take the initial pulse to be a fundamental bright soliton of the form $A(t,z=0)=A_{0}\sech{A_{0}t}$, with $A_{0}$ being the input soliton amplitude. Starting with a fundamental soliton will make the simulations particularly clear in the next section dedicated to numerical results, since there are no complications due to soliton splitting. The initial conditions for the quantum fluctuations are set by posing $U(z=0)=I$ and $V(z=0)=0$, i.e. we have an initial vacuum.

The phase-matching condition for the emitted radiation is given by $D(\omega)+\gamma A_{0}^2/2=0$, where the second term is the nonlinear soliton wavenumber shift. We solve this equation in $\omega$ to find the real roots, which gives quite an accurate prediction of where the RR appears in the spectrum. In the presence of only $\beta_{3}$ as the higher-order-dispersion coefficient, we have a single real root $\omega_{\RR}$.

We then define a RR frequency window $\mathcal W=\{k:|\omega_{k}-\omega_{\RR}|\leq\Delta\omega_{\RR}/2\}$, where $\Delta\omega_{\RR}$ is a user-defined range, typically chosen in such a way that it will contain the whole spectrum of the RR during its entire spatial evolution, taking care that the chosen range does not overlap too much with the soliton spectrum. In the particular example chosen here and in the next numerical section, this is straightforward since the RR remains in the fixed range, however in other more complicated settings, such as in supercontinuum generation and multi-soliton excitation, the frequency window should be dynamically shifted to follow the radiation during the spatial propagation.

After the above preliminary settings, we start the classical pulse propagation without the quantum fluctuations, by implementing the Strang splitting (\ref{eq:strang}). We perform half a linear step taking $\tilde{A}={\rm FFT}(A)$, multiply by the dispersion propagator $P(\omega)$ and return to the amplitude in the time domain with $A={\rm IFFT}(P(\omega)\tilde{A})$. We store the resulting mid-step field and call it $A_{\rm mid}(t)$, which will be used later in the quantum step. We then continue with the classical nonlinear full-step propagation by integrating the local nonlinear evolution over $\deltaz$ with a numerical routine of our choice (typically a fourth-order Runge--Kutta algorithm), and then apply a second linear dispersive half step to finish the Strang splitting.

We now come to the quantum step. The next thing to do is advance the matrices $U$ and $V$ in space from $z$ to $z+\Delta z$. The linear dispersive half step is performed similarly to the analogous step for the classical field: $U_{\omega}\leftarrow {\rm diag}(P(\omega))U_{\omega}$ and $V_{\omega}\leftarrow {\rm diag}(P(\omega))V_{\omega}$, where from now on (in order to avoid possible confusions) we shall use the notation $U_{\omega},V_{\omega}$ and $U_{t},V_{t}$ to indicate the Bogoliubov matrices in the frequency and time domain, respectively. We further prepare the time domain $U$ and $V$ matrices via Fourier transform unitaries $U_{t}=F^{\dagger}U_{\omega}F$ and $V_{t}=F^{\dagger}V_{\omega}F^{*}$. The complex conjugation in $F^{*}$ in the expression for $V_{t}$ is present because $V$ multiplies creation operators, and creation operators transform with the complex-conjugate mode functions.

For the full step $\Delta z$ associated to the evolution of $U$ and $V$ due to the local Kerr nonlinearity, we use the previously stored mid-step field $A_{\rm mid}(t)$ and substitute it into the BdG expressions \eqref{eq:Hlocal} and \eqref{eq:uv_closed}, which give us the $\alpha_n$, $\mu_n$ and $s_n$ coefficients. We then construct the coefficients $u_{n}$ and $v_{n}$ of the Bogoliubov propagator $e^{-iH_n\deltaz}$ of Eq. (\ref{eq:Hlocal}), and for each time row $n$ we update the $U_t$ and $V_t$ matrices with:
\begin{equation}
U_t'(n,:)=u_n U_t(n,:)+v_n\,V_t(n,:)^{*},\qquad
V_t'(n,:)=u_n V_t(n,:)+v_n\,U_t(n,:)^{*},
\label{eq:rowupdate}
\end{equation}
where we have used a typical MATLAB notation in this formula.

The last step of one QSSF propagation interval is then to go back to the frequency domain with $U_{\omega}=FU_{t}F^{\dagger}$, $V_{\omega}=FV_{t}F^{T}$ (a transpose appears in the latter expression due to $(F^{*})^{-1}=F^{T}$ specifically for the Fourier transform unitary), and then perform again a half-step linear propagation with $U_{\omega}\leftarrow {\rm diag}(P(\omega))U_{\omega}$ and $V_{\omega}\leftarrow {\rm diag}(P(\omega))V_{\omega}$. This completes both the classical and quantum advancement of the fields from $z$ to $z+\Delta z$.

At any desired propagation distance, the matrices $U(z)$ and $V(z)$ are restricted to the chosen spectral window and used to calculate the second moments through Eq.~(\ref{eq:NM_UV}). These second moments give the covariance matrix $V_q$ through Eq.~(\ref{eq:Vq_NM}). The symplectic eigenvalues, obtained from the positive moduli of the spectrum of $i\Omega V_q$, are then used to calculate the von Neumann entropy, R\'enyi-2 entropy, purity, symplectic occupations, and effective number of active Williamson modes via Eqs.~\eqref{eq:SvN}--\eqref{eq:Keff}.

\section{Numerical results and entanglement evolution}

To illustrate the capabilities of the QSSF algorithm, we consider the propagation of a fundamental NLSE soliton in the presence of third-order dispersion. The initial condition is taken to be $A(z=0,t)=A_0\,\mathrm{sech}(A_0 t)$, with $A_{0}=3$, and vacuum fluctuations in all FFT modes, corresponding to $U(0)=I$ and $V(0)=0$.
Throughout the simulation the classical field is propagated (total propagation is $L=7$ dispersion lengths, $\Delta z=10^{-3}$, temporal window $T=150$, number of temporal sampling points $N_{t}=2048$) using the symmetric Strang splitting while the quantum fluctuations are evolved using the commutator-preserving Bogoliubov propagation described in the previous sections.

With $\gamma=1$, the third-order dispersion coefficient $\beta_{3}=0.05$ generates a phase-matched resonant-radiation (RR) band at a frequency $\omega_{\RR}$ determined by the phase-matching condition
$D(\omega_{\RR})+\gamma A_0^2/2=0$, giving $\omega_{\RR}\simeq-10.78$.
A frequency window $\mathcal W$ spanning the range $\omega\in[-16,-7]$ is therefore selected around $\omega_{\RR}$ (corresponding to $214$ FFT bins). The reduced Gaussian state associated with this frequency interval is then dynamically constructed throughout the propagation using the Bogoliubov matrices $(U,V)$.

The numerical accuracy of the QSSF algorithm is demonstrated by both the classical and quantum diagnostics. At the end of propagation the relative energy error remains $|E(L)-E(0)|/E(0)\simeq 5.24\times10^{-6}$, where $E(z)\equiv\int|A(z,t)|^{2}dt$, while the symplecticity indicators remain extremely small, $\varepsilon_{1,2}\lesssim10^{-11}$,
showing that the Bogoliubov propagation preserves the canonical commutation relations to essentially machine precision. 

As an additional validation test, we explicitly reconstructed the covariance matrix of the complete simulated spectrum and computed its full Gaussian von Neumann entropy from the corresponding symplectic eigenvalues. Since the BdG evolution is globally unitary and the input fluctuation state is vacuum, the full Gaussian state should remain pure throughout propagation. Numerically we find
$S_{\rm total}(z)\lesssim1.3\times10^{-10}$
for all propagation distances at which this diagnostic is evaluated, confirming the internal consistency of the covariance-matrix reconstruction, symplectic-spectrum calculation, and entropy evaluation.

\begin{figure}[t]
\centering
\includegraphics[width=0.5\linewidth]{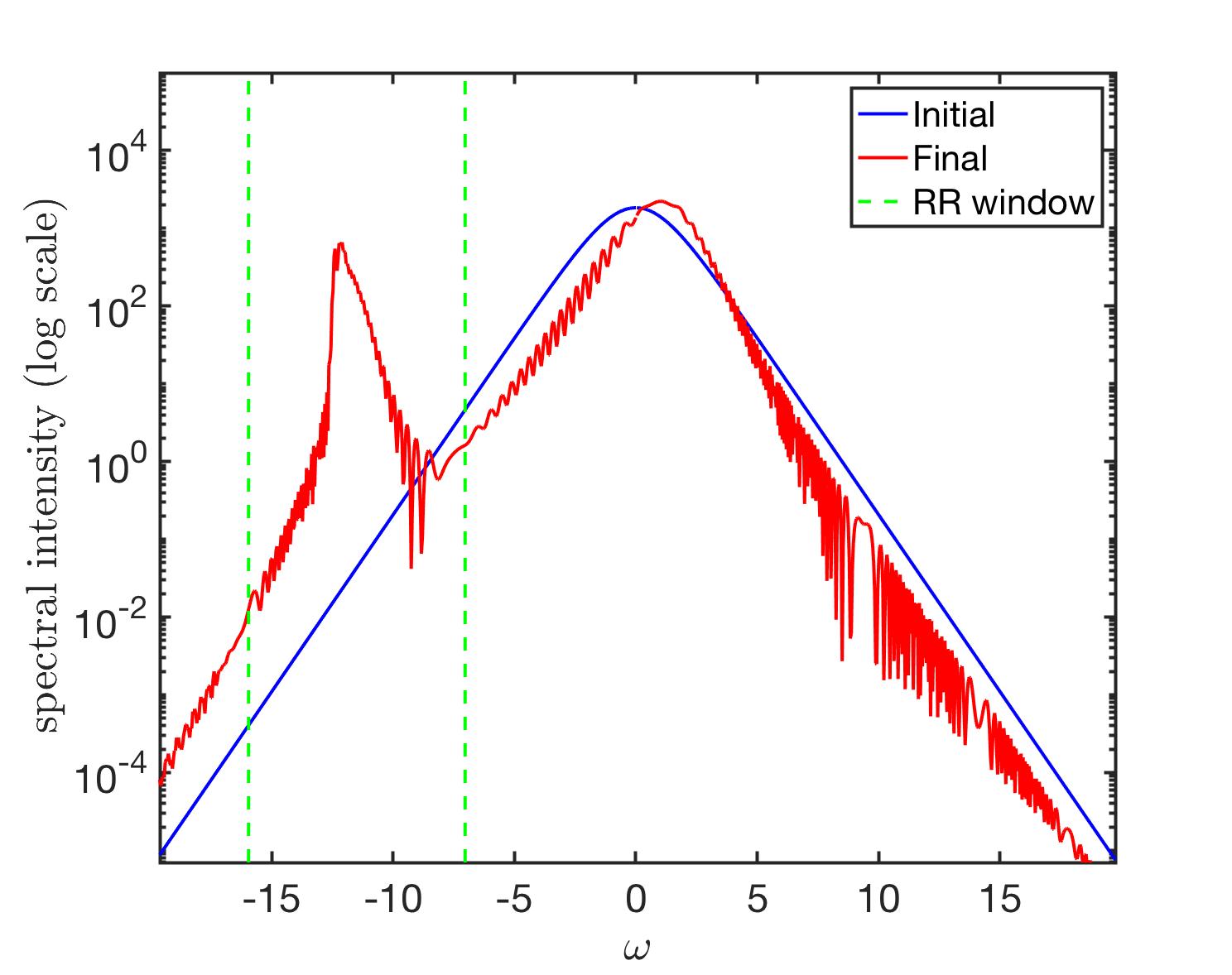}
\caption{
Spectral intensity $|\tilde A(\omega,z)|^2$ (in log scale) for the initial soliton (blue solid line, $z=0$) and the final field (red solid line, $z=7$). The resonant-radiation band emerges near the phase-matched frequency $\omega_{\RR}\simeq=-10.78$ and progressively grows in amplitude and undergoes a slight shift towards more negative detunings, consistent with soliton recoil. Vertical green dashed lines indicate the analysis window used for the reduced Gaussian state. The third-order dispersion parameter is $\beta_{3}=0.05$, and soliton amplitude is $A_{0}=3$.
}
\label{fig:spectrum_map}
\end{figure}

Figure~\ref{fig:spectrum_map} shows the evolution of the optical spectrum during propagation. The RR band appears at the phase-matched frequency predicted from the analytical condition and grows monotonically with distance. The dashed lines identify the spectral window used for the reduced-state analysis.

\begin{figure}[t]
\centering
\includegraphics[width=0.5\linewidth]{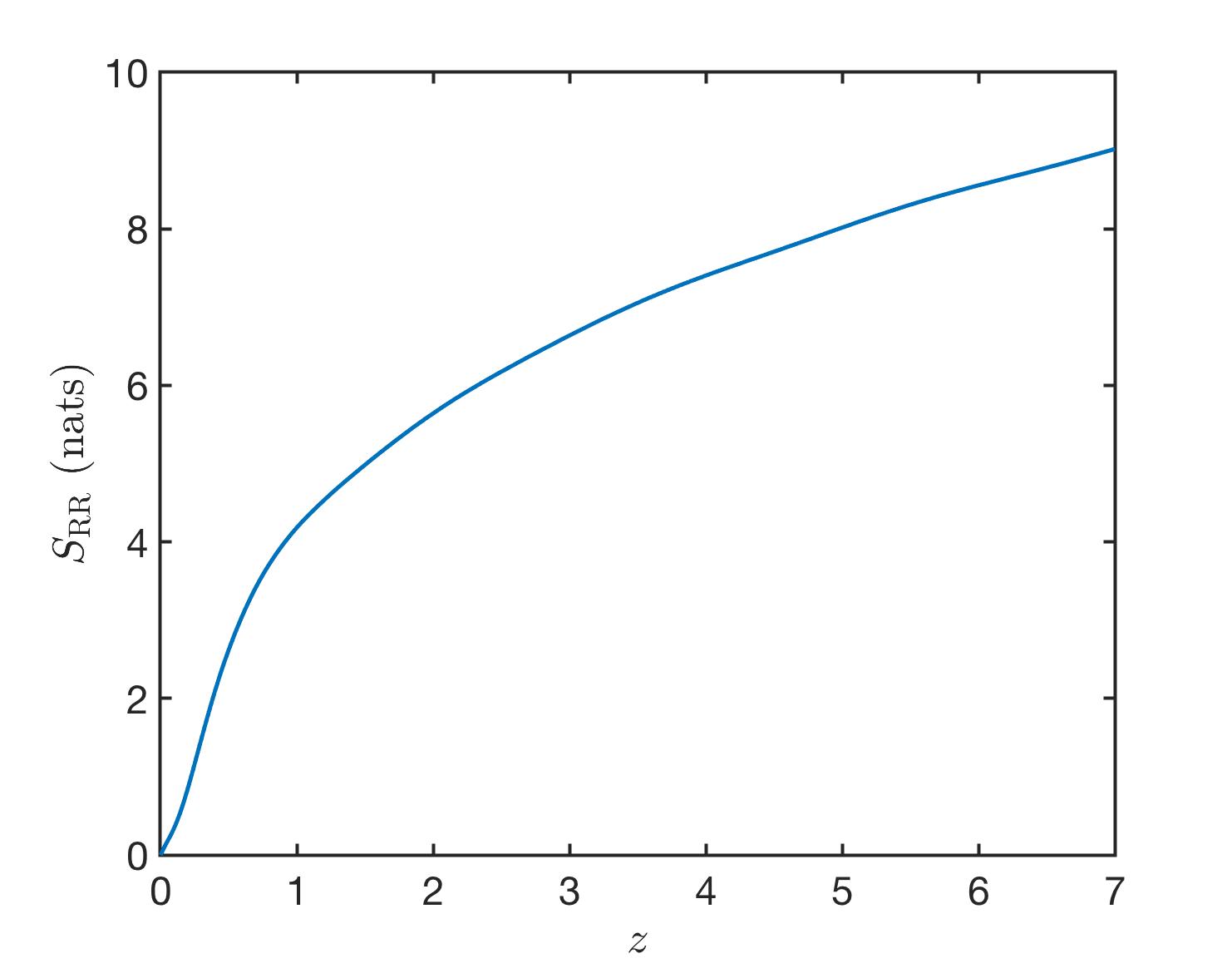}
\caption{
Von Neumann entropy (in nats) of the RR subsystem versus propagation distance. Since the global BdG state remains pure, this quantity is also the entanglement entropy between the RR spectral window and its complement. All parameters are as in Fig. \ref{fig:spectrum_map}.
}
\label{fig:entropy}
\end{figure}

The central quantity computed by the QSSF is the von Neumann entropy of the RR subsystem, $S_{\RR}(z)\equiv S_{\rm vN}(\rho_{\RR})$,
which in the present lossless Gaussian setting is exactly equal to the entanglement entropy between the RR frequency window and the remainder of the spectrum. Figure~\ref{fig:entropy} shows the evolution of $S_{\RR}(z)$ with propagation length. Starting from vacuum fluctuations, $S_{\RR}(0)=0$,
the entropy increases monotonically throughout propagation and reaches a value of $S_{\RR}(L)=9.02$ nats at the output. This behaviour demonstrates the continuous generation of spectral entanglement during RR emission.

\begin{figure}[t]
\centering
\includegraphics[width=0.5\linewidth]{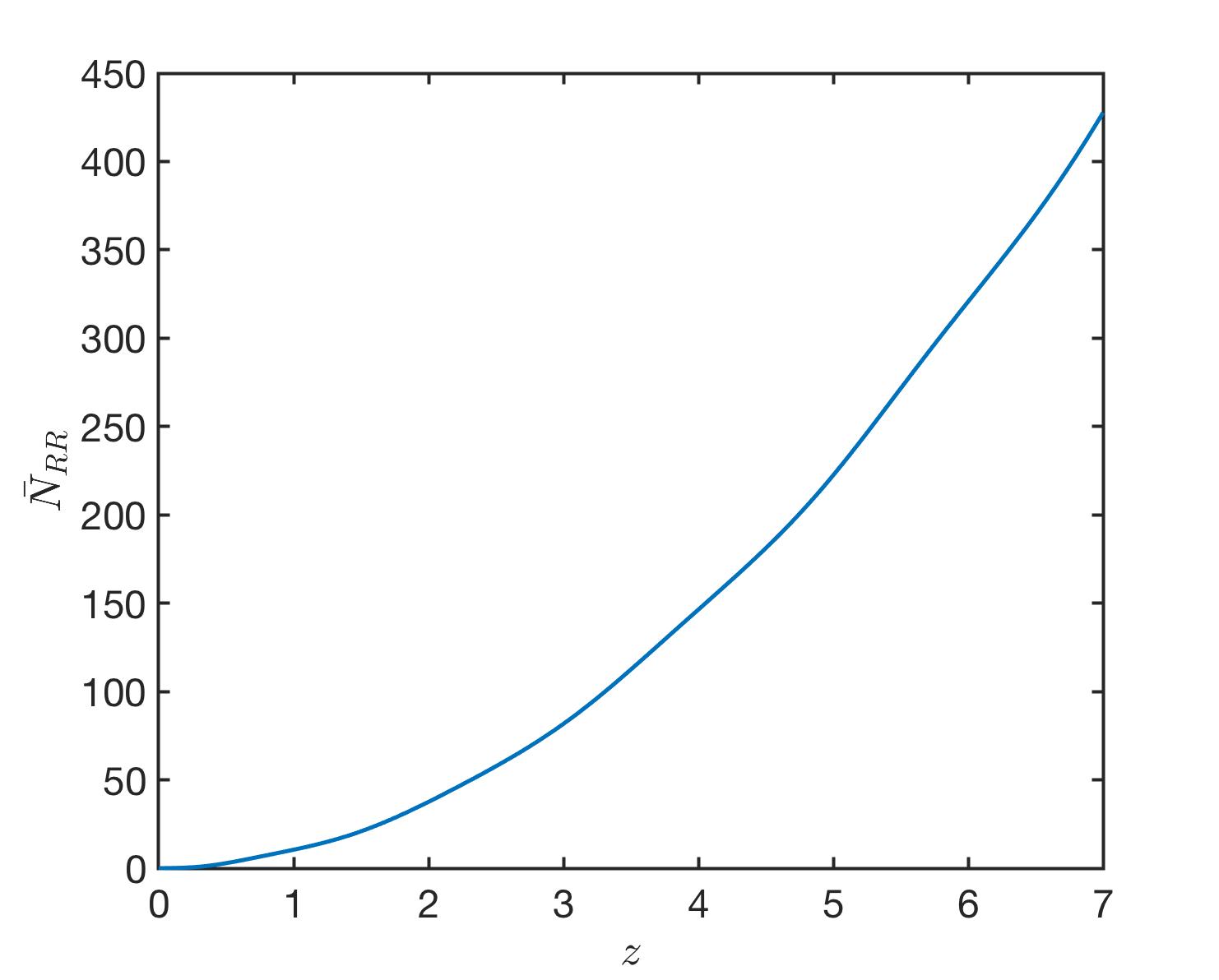}
\caption{
Evolution of the total fluctuation population inside the RR spectral window. The continuous increase of $\bar N_{\RR}$ reflects the amplification and generation of quantum fluctuations associated with RR formation. All parameters are as in Fig. \ref{fig:spectrum_map}.
}
\label{fig:photon_number}
\end{figure}

The growth of entanglement is accompanied by the accumulation of quantum fluctuations inside the RR window $\mathcal W$. Figure~\ref{fig:photon_number} reports the total photon population
$\bar N_{\RR}\equiv\Tr(N_{\mathcal W})=\sum_{k\in \mathcal W}\langle a_{k}^{\dagger}a_{k} \rangle$, which reaches
$\bar N_{\RR}(L)=427.63$.
The growth of \(\bar N_{\RR}\) shows that the selected RR band accumulates a substantial fluctuation population during propagation.

\begin{figure}[t]
\centering
\includegraphics[width=0.5\linewidth]{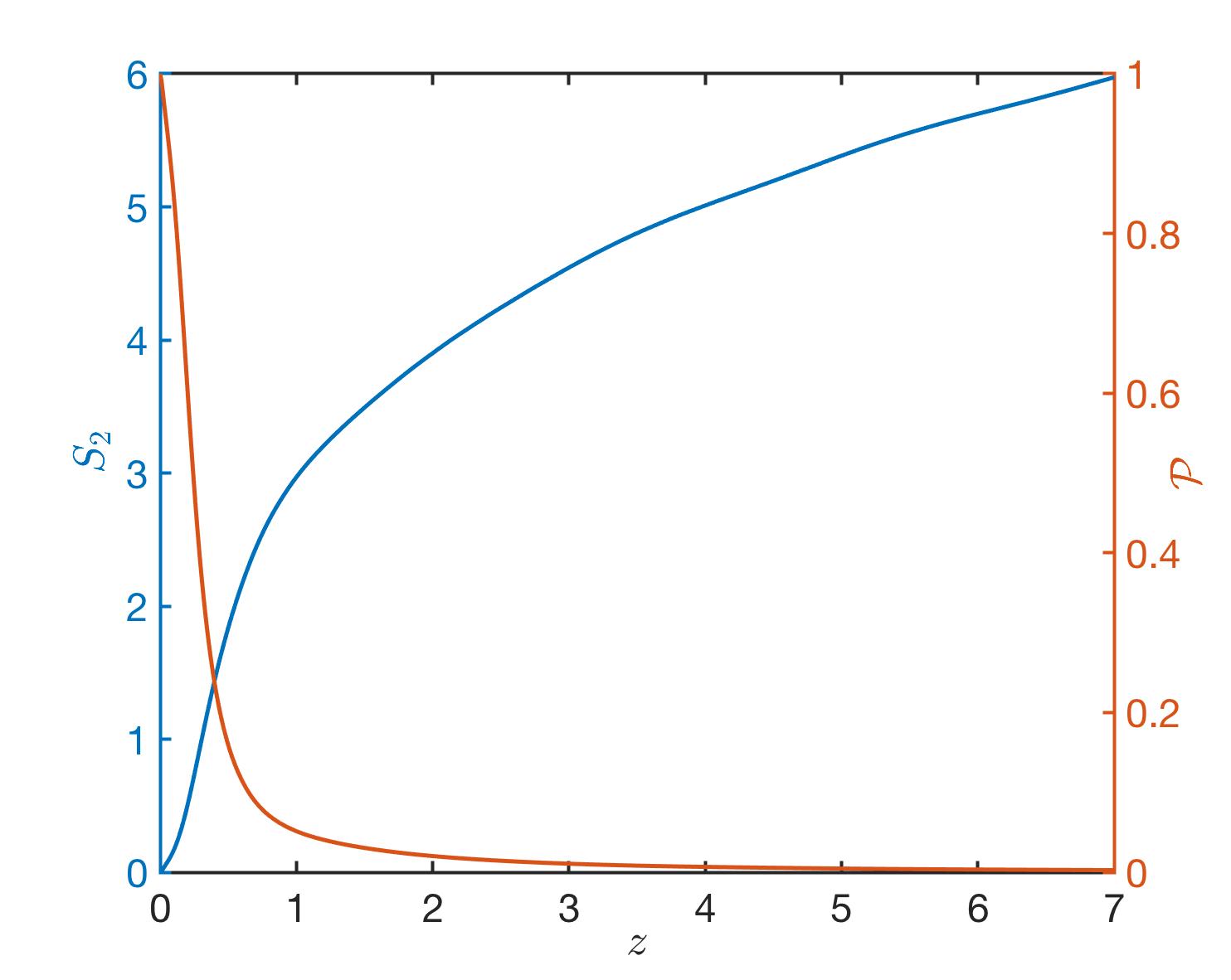}
\caption{
R\'enyi-2 entropy (blue solid line, axis on the left) and purity (red solid line, axis on the right) of the RR reduced state. The loss of purity is entirely due to entanglement, since the global Gaussian state remains pure within the BdG approximation. All parameters are as in Fig. \ref{fig:spectrum_map}.
}
\label{fig:purity}
\end{figure}

The reduced RR state also becomes increasingly mixed. Figure~\ref{fig:purity} presents the R\'enyi-2 entropy and purity, see Eqs. (\ref{eq:S2}).
The purity decreases from unity at the input to
$\mathcal P(L)\simeq 2.54\times10^{-3}$ at the end of the propagation, demonstrating once again strong entanglement between the RR subsystem and the rest of the spectrum.

\begin{figure}[t]
\centering
\includegraphics[width=0.5\linewidth]{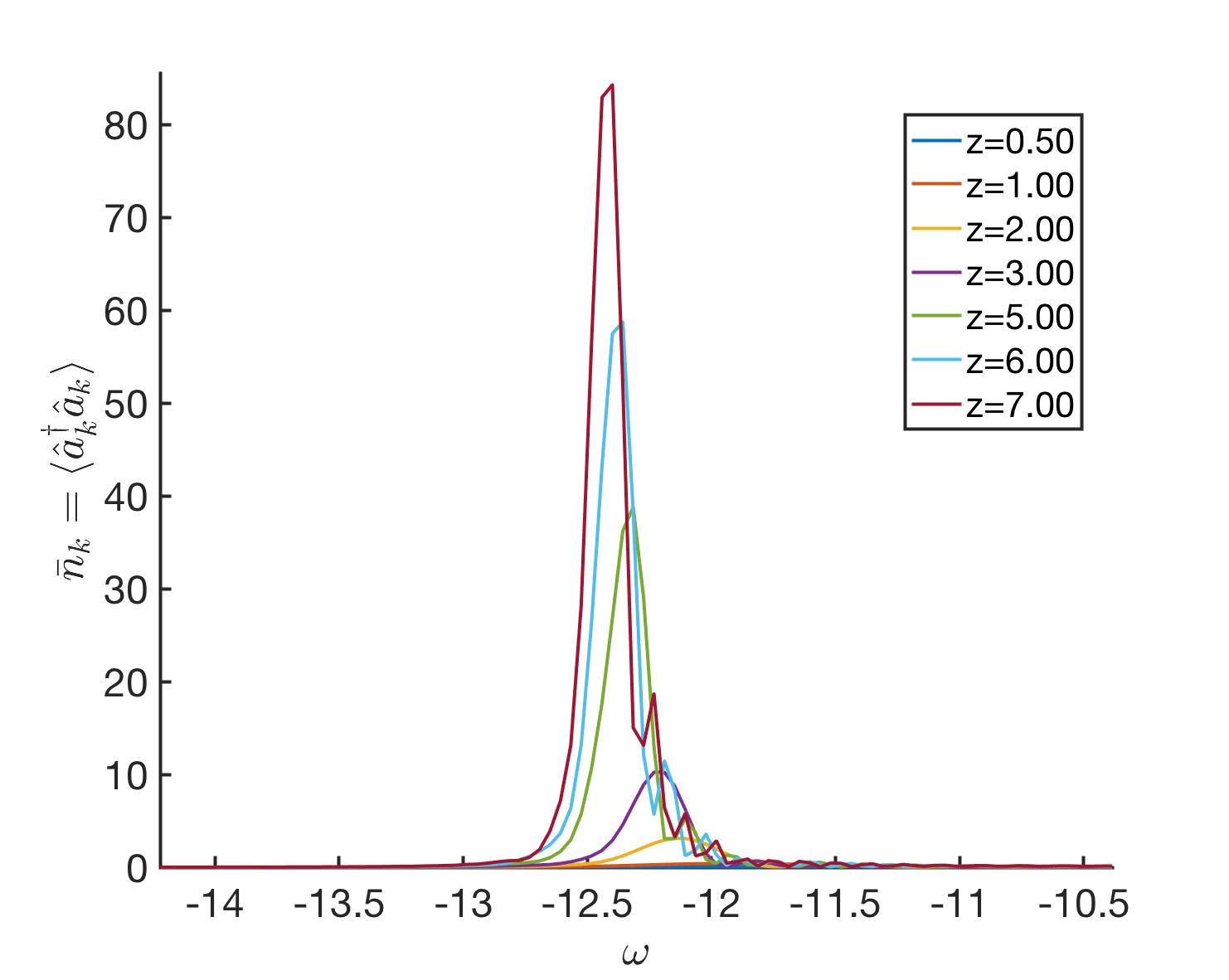}
\caption{
Mean occupations of individual FFT bins inside the resonant-radiation window at several propagation distances (different colours for different propagation distances, see the legend). Although many frequency bins become populated, this is a basis-dependent description that does not reveal the intrinsic structure of the Gaussian state. All parameters are as in Fig. \ref{fig:spectrum_map}.
}
\label{fig:fft_modes}
\end{figure}

A major advantage of the QSSF framework is the ability to distinguish between occupations of FFT modes and occupations of the intrinsic symplectic modes of the reduced Gaussian state. Figure~\ref{fig:fft_modes} shows the mean occupations
$\bar n_k=\langle \hat a_k^\dagger\hat a_k\rangle$ of the FFT bins within the RR window at several propagation distances. The population is distributed across many frequency bins and broadens as propagation proceeds.

\begin{figure}[t]
\centering
\includegraphics[width=0.5\linewidth]{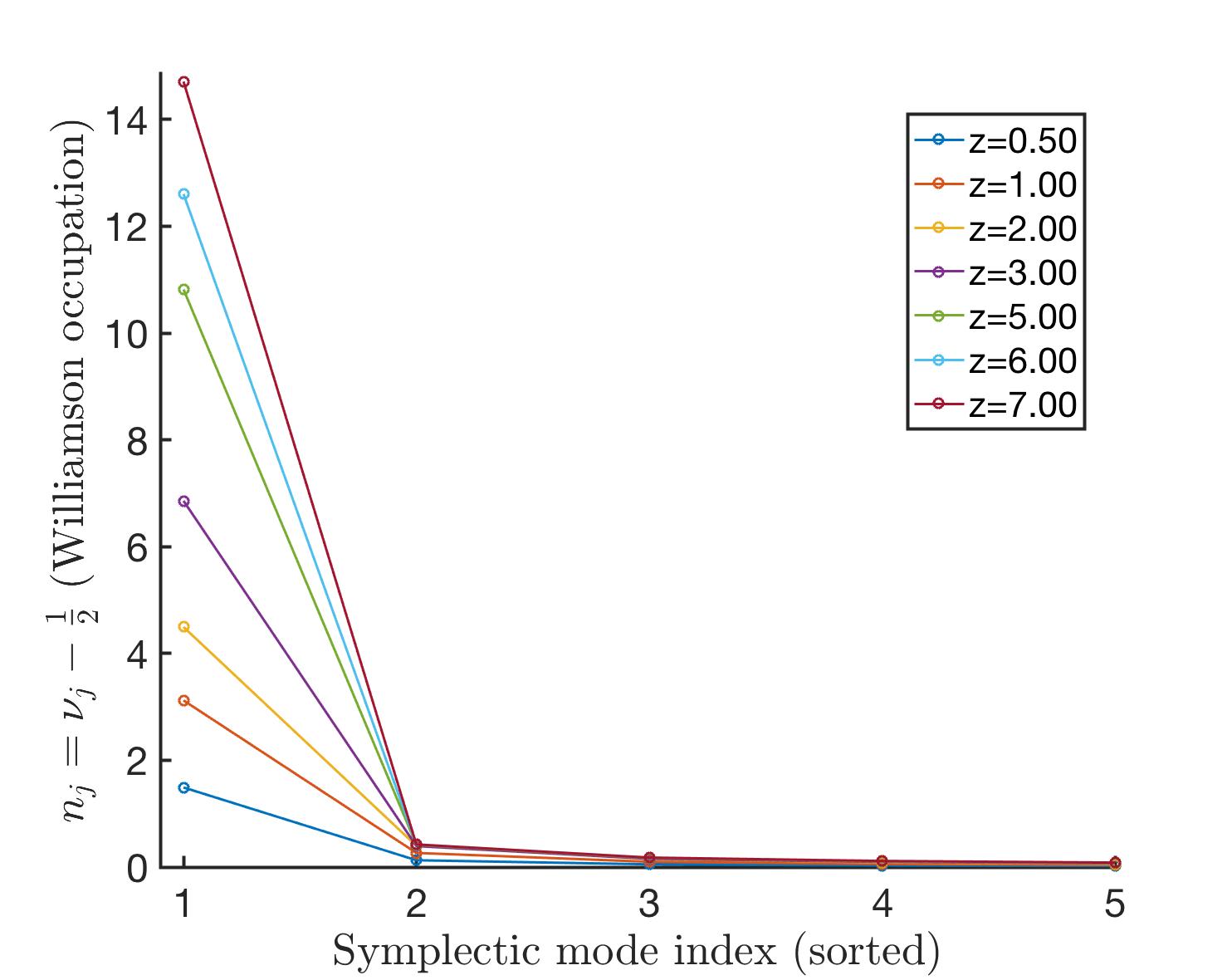}
\caption{
Sorted symplectic occupations of the RR reduced state at different propagation distances (different colours for different propagation distances, see the legend). The rapid decay of the spectrum reveals the emergence of a dominant Williamson mode. The plot has been zoomed to clearly show the rapidity of formation of dominant Williamson modes; in reality the mode index on the horizontal axis goes from 1 to 214, which is the total number of symplectic modes in the RR window. All parameters are as in Fig. \ref{fig:spectrum_map}.
}
\label{fig:symplectic_modes}
\end{figure}

The Williamson decomposition provides a basis-independent description. Figure~\ref{fig:symplectic_modes} displays the sorted symplectic occupations
$n_j=\nu_j-\frac12$ for various propagation distances.
A striking hierarchy is observed: the occupation spectrum rapidly decays, leaving one (or in the general case a few) dominant Williamson mode(s) and a large number of weakly populated modes.

\begin{figure}[t]
\centering
\includegraphics[width=0.5\linewidth]{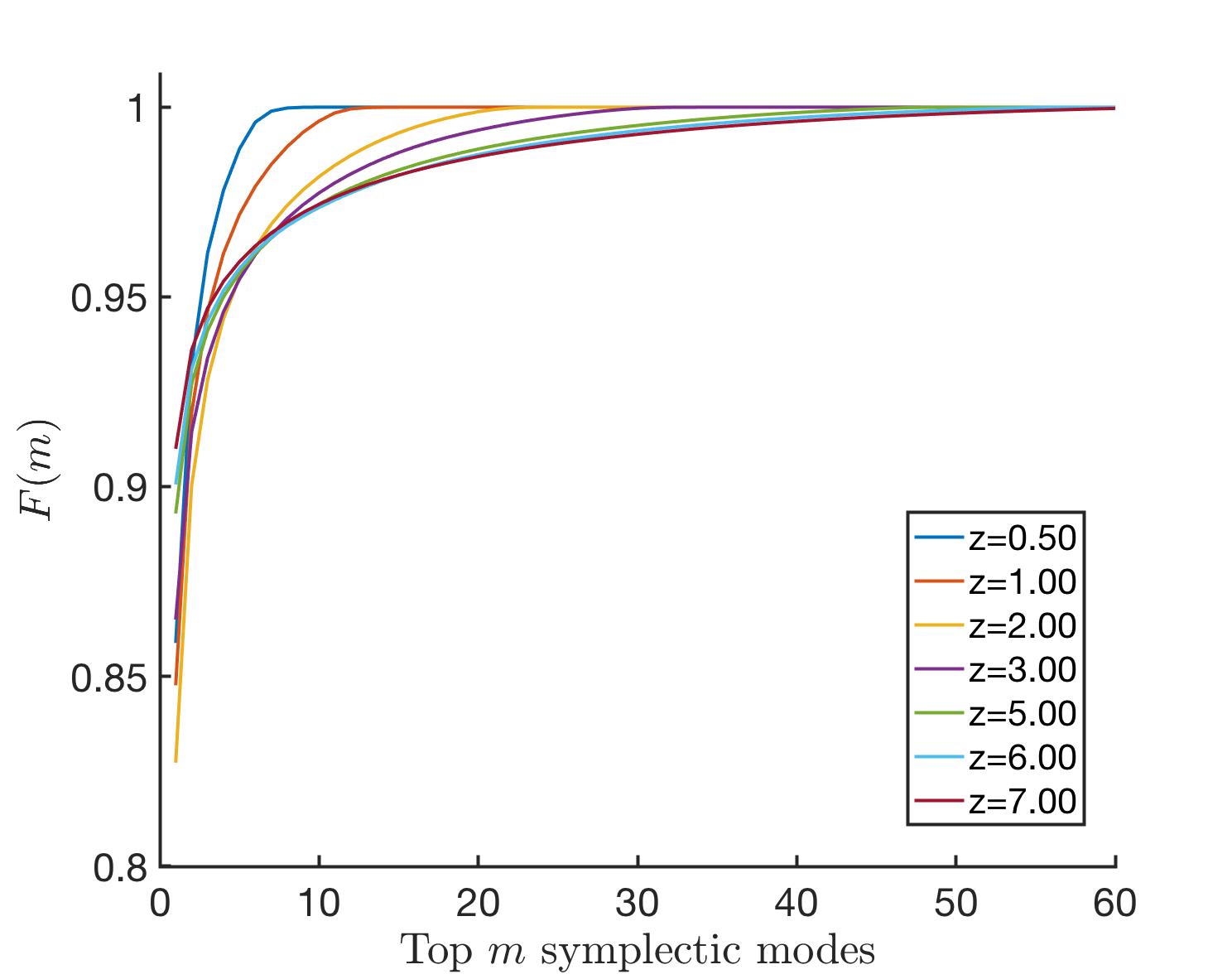}
\caption{
Cumulative weight $F(m)$ of the sorted symplectic occupations (different colours for different propagation distances, see the legend). The first Williamson mode rapidly captures the overwhelming majority of the total occupation, demonstrating that the RR subsystem becomes effectively single-mode in symplectic space. All parameters are as in Fig. \ref{fig:spectrum_map}.}
\label{fig:cumulative}
\end{figure}

This dimensional reduction becomes particularly clear in Figure~\ref{fig:cumulative}. The cumulative occupation weight
\[
F(m)=
\frac{\sum_{j=1}^{m} n_j}
{\sum_{j=1}^{n} n_j}
\]
approaches unity after only a few Williamson modes. At large propagation distances the leading Williamson mode carries approximately \(92\%\) of the total symplectic occupation in the RR window. Consistent with this observation, the effective mode number [see Eq.~(\ref{eq:Keff})] reaches only $K_{\rm eff}=1.206$ despite the fact that the RR subsystem contains 214 FFT bins. This demonstrates a remarkable dimensional reduction: although the RR occupies an extended spectral region, its quantum state is essentially dominated by a single (or in general very few) collective Gaussian degree(s) of freedom.

\begin{figure}[t]
\centering
\includegraphics[width=0.8\linewidth]{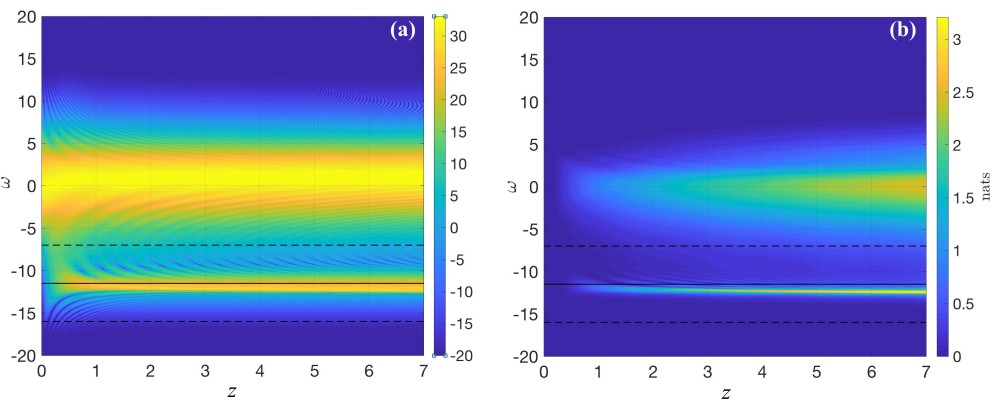}
\caption{
(a) Evolution of the input soliton spectrum with $z$ (intensity is in arbitrary log scale); the soliton and the emission of the RR are clearly visible. The black solid line is the predicted phase-matched frequency of the RR, while the black dashed lines indicate the RR window considered in the entropy calculation. (b) Single-bin entropy map $S_k(\omega,z)$. Bright regions identify FFT bins whose single-mode reduced states become strongly mixed, equivalently entangled with the remaining modes in the lossless pure BdG setting. The map displays both the broad contribution of the soliton-core region and a narrow feature associated with the selected RR band. All parameters are as in Fig. \ref{fig:spectrum_map}.
}
\label{fig:entanglement_map}
\end{figure}

Finally, Figure~\ref{fig:entanglement_map}(a) presents the traditional soliton spectral evolution along the fiber, while Fig. \ref{fig:entanglement_map}(b) shows the spectral map of the single-bin reduced entropy $S_k(z)$, calculated independently for every FFT frequency bin. In the lossless pure BdG setting, this quantity measures the entanglement of an individual FFT mode with the remainder of the field, although it is a basis-dependent single-mode diagnostic. The figure provides a direct visualization of where quantum correlations are generated throughout the spectrum. The map shows both broad soliton-core contributions and a narrow correlated feature inside the selected RR window. In this sense, the entropy map may be viewed as a quantum-information analogue of the standard spectral-evolution diagram of Fig. \ref{fig:entanglement_map}(a), while keeping in mind that the single-bin entropies are not additive over frequency bins.

Overall, the simulations demonstrate that the RR generation is accompanied by substantial production of Gaussian entanglement, while simultaneously exhibiting an unexpected and pronounced compression into a nearly single Williamson mode. The QSSF algorithm therefore reveals information that is inaccessible to conventional intensity- or coherence-based analyses and provides a direct route to quantifying the quantum structure of nonlinear frequency-conversion processes.

We finally comment on the computational efficiency of the algorithm. The principal computational cost of QSSF arises from the propagation of the Bogoliubov matrices $U$ and $V$, which are dense $N_t\times N_t$ matrices. Each propagation step requires matrix Fourier transforms and time-local Bogoliubov updates, leading to an overall propagation cost approximately proportional to $N_t^2\log N_t$ per longitudinal step. If $N_z=L/\Delta z$ is the number of propagation steps, the dominant propagation cost scales as $O(N_zN_t^2\log N_t)$, while the memory cost scales as $O(N_t^2)$.

For a reduced spectral window containing $n$ modes, the construction of the restricted second moments $N_{\mathcal W}=V_{\mathcal W}^{*}V_{\mathcal W}^{T}$ and $M_{\mathcal W}=U_{\mathcal W}V_{\mathcal W}^{T}$ scales as $O(n^2N_t)$. Once these moments are available, assembling the $2n\times2n$ covariance matrix is an $O(n^2)$ operation, while evaluating the symplectic spectrum from the eigenvalues of $i\Omega V_q$ scales as $O(n^3)$. Thus the entropy diagnostic for one window scales as $O(n^2N_t+n^3)$. In typical applications the reduced window is much smaller than the full simulation bandwidth ($n\ll N_t$), so the dominant cost is the propagation of $(U,V)$ rather than the entropy evaluation itself. Full symplecticity checks involving products such as $UU^\dagger$ and $UV^T$ scale as $O(N_t^3)$ and are therefore best performed only at selected checkpoints in production runs.

For the example presented in this work ($N_t=2048$, $L=7$, $\Delta z=10^{-3}$), corresponding to $7000$ propagation steps and a resonant-radiation window containing $214$ FFT bins, the simulation required approximately $5\times10^3$\,s on a standard workstation while maintaining energy conservation at the level of $10^{-6}$ and symplecticity errors below $10^{-11}$. The algorithm is therefore computationally demanding but remains practical for high-resolution studies of quantum correlations in nonlinear waveguides.

\section{Conclusions and future work}

In conclusion, we have introduced the Quantum Split-Step Fourier algorithm, a numerical framework that combines conventional split-step propagation of the nonlinear Schr\"odinger equation with a symplectic Bogoliubov evolution of quantum fluctuations. The method propagates the Bogoliubov matrices $(U,V)$ together with the classical optical field, preserving the canonical commutation relations and thereby maintaining a physically consistent Gaussian quantum evolution.

A key advantage of the method is that the complete Gaussian state of any chosen spectral subsystem can be reconstructed directly from the propagated Bogoliubov matrices. This enables the dynamic calculation of covariance matrices, symplectic spectra, von Neumann entropies, R\'enyi entropies, purities, mode occupations, and effective modal dimensionalities during nonlinear pulse propagation. In contrast with traditional coherence studies based on ensemble averages, QSSF provides a direct information-theoretic characterization of quantum correlations and entanglement.

Application of the algorithm to soliton-driven resonant radiation reveals several generic features. The resonant radiation subsystem acquires a steadily increasing entanglement entropy as propagation proceeds, while its purity decreases due to entanglement with the remainder of the spectrum. Despite the large number of available FFT modes, the reduced state is found to be dominated by only a few Williamson modes, indicating a pronounced low-rank structure in the underlying Gaussian correlations. Contour maps of single-bin entropy further show how entanglement is generated spectrally and provide a new visualization tool complementary to conventional supercontinuum diagnostics.

The present implementation represents only a first step. Future extensions may incorporate generalized-NLSE effects such as Raman scattering, self-steepening, frequency-dependent nonlinearities, linear loss, distributed gain, plasma contributions in gas-filled waveguides, and multimode propagation. Beyond Gaussian states, it would also be interesting to explore hybrid approaches capable of describing strongly non-Gaussian quantum fluctuations while retaining the computational efficiency of the split-step framework.

Because the algorithm is formulated directly in terms of physically measurable spectral windows, QSSF provides a practical bridge between nonlinear waveguide dynamics and continuous-variable quantum information. We expect it to become a useful computational tool for the analysis and design of quantum frequency converters, supercontinuum sources, resonant radiation platforms, multimode squeezed light generators, and future integrated photonic architectures exploiting spectrally distributed entanglement.

\section{Acknowledgements}
The author would like to acknowledge useful discussions with Dr Chris Brahms, Prof John Travers and Prof Alessandro Fedrizzi (all from Heriot-Watt University, Edinburgh, UK).

\end{document}